\documentclass[12pt]{article}
\usepackage{epsf}
\usepackage{graphicx}

\begin{document}
\title
{Gravitons as super-strong interacting particles, and low-energy
quantum gravity}
\author
{Michael A. Ivanov \\
Physics Dept.,\\
Belarus State University of Informatics and Radioelectronics, \\
6 P. Brovka Street,  BY 220027, Minsk, Republic of Belarus.\\
E-mail: ivanovma@gw.bsuir.unibel.by.}

\maketitle

\begin{abstract}
It is shown by the author that if gravitons are super-strong
interacting particles and the low-temperature graviton background
exists, the basic cosmological conjecture about the Dopplerian
nature of redshifts may be false. In this case, a full magnitude
of cosmological redshift would be caused by interactions of
photons with gravitons. A new dimensional constant which
characterizes one act of interaction is introduced and estimated.
Non-forehead collisions with gravitons will lead to a very
specific additional relaxation of any photonic flux. It gives a
possibility of another interpretation of supernovae 1a data -
without any kinematics. Of course, all of these facts may
implicate a necessity to change the standard cosmological
paradigm. Some features of a new paradigm are discussed here, too.
\par A quantum mechanism of classical gravity based on an
existence of this sea of gravitons is described for the Newtonian
limit. This mechanism needs graviton pairing and "an atomic
structure" of matter for working it, and leads to the time
asymmetry. If the considered quantum mechanism of classical
gravity is realized in the nature, then an existence of black
holes contradicts to Einstein's equivalence principle. It is shown
that in this approach the two fundamental constants - Hubble's and
Newton's ones - should be connected between themselves. The
theoretical value of the Hubble constant is computed. In this
approach, every massive body would be decelerated due to
collisions with gravitons that may be connected with the Pioneer
10 anomaly. It is shown that the predicted and observed values of
deceleration are in good agreement. Some unsolved problems are
discussed, so as possibilities to verify some conjectures in
laser-based experiments.
\end{abstract}
\section[1]{Introduction }
By a full coincidence of the forms of Coulomb's and Newton's laws,
which describe an interaction of electric charges and a
gravitational interaction of bodies, we see a dramatic difference
in developing the pictures of these interactions on a quantum
level. Constructed with pillars on multiple experiments QED is one
of the most exact physical theories and an archetype for imitation
by creation of new models. While the attempts to quantize the
remarkable in its logical beauty theory of general relativity,
which describes gravitation on a classical level so fully and
delicately, (see review \cite{33}) not only have not a hit until
today but gave a specific side psychological effect - there exists
conceptualization that quantum gravity may be described only by
some sophisticated theory. An opinion is commonly accepted, too,
that quantum gravity should manifest itself only on the Planck
scale of energies, i.e. it is a high-energy phenomenon. The value
of the Planck energy $\sim 10^{19}$ GeV has been got from
dimensional reasonings. Still one wide-spread opinion is that we
know a mechanism of gravity (bodies are exchanging with gravitons
of spin 2) but cannot correctly describe it. If an apple from the
legend about Newton's afflatus can imagine all these
complications, it would hesitate to fall to the ground so
artlessly as it is accepted among the apples. \par Perhaps,
physicists would be able to refuse easier the preconceived
stereotypes which balk - as it seems to me - to go ahead in
understanding quantum gravity if experiments or observations would
give more essential meat for reasonings. But in this area, at
least up to recent years, there was observed nothing that may
serve if not Ariadne's clew but such a simple physical
contradiction that an aspiration to overcome it would advantage
introduction of new ideas and revision of the "inviolable".

\par In a few last years, the situation has been abruptly changed. I
enumerate those discoveries and observations which may force, in
my opinion, the ice to break up. \par 1. In 1998, Anderson's team
reported about the discovery of anomalous acceleration of NASA's
probes Pioneer 10/11 \cite{1}; this effect is not embedded in a
frame of the general relativity, and its magnitude is somehow
equal to $\sim Hc$, where $H$ is the Hubble constant, $c$ is the
light velocity. \par 2. In the same 1998, two teams of
astrophysicists, which were collecting supernovae 1a data with the
aim to specificate parameters of cosmological expansion, reported
about dimming remote supernovae \cite{2,3}; the one would be
explained on a basis of the Doppler effect if at present epoch the
universe expands with acceleration. This explanation needs an
introduction of some "dark energy" which is unknown from any
laboratory experiments. \par 3. In January 2002, Nesvizhevsky's
team  reported about discovery of quantum states of ultra-cold
neutrons in the Earth's gravitational field \cite{4}. Observed
energies of levels (it means that and their differences too) in
full agreement with quantum-mechanical calculations turned out to
be equal to $\sim 10^{-12}$ eV. The formula for energy levels had
been found still by Bohr and Sommerfeld. If transitions between
these levels are accompanied with irradiation of gravitons then
energies of irradiated gravitons should have the same order - but
it is of 40 orders lesser than the Planck energy by which one
waits quantum manifestations of gravity. \par The first of these
discoveries obliges to muse about the borders of applicability of
the general relativity, the third - about that quantum gravity
would be a high-energy phenomenon. It seems that the second
discovery is far from quantum gravity but it obliges us to look at
the traditional interpretation of the nature of cosmological
redshift critically. An introduction into consideration of an
alternative model of redshifts \cite{5} which is based on a
conjecture about an existence of the graviton background gives us
odds to see in the effect of dimming supernovae an additional
manifestation of low-energy quantum gravity. Under the definite
conditions, an effective temperature of the background may be the
same one as a temperature of the cosmic microwave background, with
an average graviton energy of the order $\sim 10^{-3}$ eV. \par In
this chapter, the main results of author's research in this
approach are described. Starting from a statistical model of the
graviton background with low temperature, it is shown - under the
very important condition that gravitons are super-strong
interacting particles - that if a redshift would be a quantum
gravitational effect then one can get from its magnitude an
estimate of a new dimensional constant characterizing a single act
of interaction in this model. It is possible to calculate
theoretically a dependence of a light flux relaxation on a
redshift value, and this dependence fits supernova observational
data very well at least for $z < 0.5$. Further it is possible to
find a pressure of single gravitons of the background which acts
on any pair of bodies due to screening the graviton background
with the bodies \cite{6}. It turns out that the pressure is huge
(a corresponding force is $\sim 1000$ times stronger than the
Newtonian attraction) but it is compensated with a pressure of
gravitons which are re-scattered by the bodies. The Newtonian
attraction arises if a part of gravitons of the background forms
pairs which are destructed by interaction with bodies. It is
interesting that both Hubble's and Newton's constants may be
computed in this approach with the ones being connected between
themselves. It allows us to get a theoretical estimate of the
Hubble constant. An unexpected feature of this mechanism of
gravity is a necessity of "an atomic structure" of matter - the
mechanism doesn't work without the one.
\par Collisions with gravitons should also call forth a deceleration
of massive bodies of order $\sim Hc$ - namely the same as of
NASA's probes. But at present stage it turns out unclear why such
the deceleration has {\it not} been observed for planets. The
situation reminds by something of the one that took place in
physics before the creation of quantum mechanics when a motion of
electrons should, as it seemed by canons of classical physics,
lead to their fall to a nucleus. \par Because the very unexpected
hypothesis founded into the basis of this approach is a
super-strong character of gravitational interaction on a quantum
level, I would like to explain my motivation which conduced namely
to such the choice.  Learning symmetries of the quantum
two-component composite system \cite{7}, I have found that its
discrete symmetries in an 8-space may be interpreted by an
observer from a 4-dimensional world as the exact global symmetries
of the standard model of particle physics if internal coordinates
of the system (the composite fermion) are rigidly fixed. This
conclusion was hard for me and took much enough time. But to
ensure almost full fixation of components of the system, an
interaction connecting them should be very strong. Because of it,
when a choice arise - an amount of gravitons or an intensity of
the interaction, - I have remembered this overpassed earlier
barrier and chose namely the super-strong interaction. Without
this property, the graviton background would not be in the
thermodynamical equilibrium with the cosmic microwave background
that could entail big difficulties in the model.
\par So, in this approach we deal with the following small quantum
effects of low-energy gravity: redshifts, its analog - a
deceleration of massive bodies, and an additional relaxation of
any light flux. The Newtonian attraction turns out to be the main
statistical effect, with bodies themselves being not sources of
gravitons - only correlational properties of {\it in} and {\it
out} fluxes of gravitons in their neighbourhood are changed due to
an interaction with bodies. There does still not exist a full and
closed theory in this approach, but even the initial researches in
this direction show that in this case quantum gravity cannot be
described separately of other interactions, and also manifest the
boundaries of applicability of a geometrical language in gravity.

\section[2]{Passing photons through the graviton background \cite{5,5a}}

Let us introduce the hypothesis, which is considered in this
approach as independent from the standard cosmological model:
there exists the isotropic graviton background. Photon scattering
is possible on gravitons $\gamma + h \to \gamma + h,$ where
$\gamma $ is a photon and $h$ is a graviton, if one of the
gravitons is virtual. The energy-momentum conservation law
prohibits energy transfer to free gravitons. Due to forehead
collisions with gravitons, an energy of any photon should decrease
when it passes through the sea of gravitons.
\par
From another side, none-forehead collisions of photons with
gravitons of the background will lead to an additional relaxation
of a photon flux, caused by transmission of a momentum transversal
component to some photons. It will lead to an additional dimming
of any remote objects, and may be connected with supernova
dimming.
\par We deal here with the
uniform non-expanding universe with the Euclidean space, and there
are not any cosmological kinematic effects in this model.
\subsection[2.1]{Forehead collisions with gravitons: an alternative
explanations of the redshift nature} We shall take into account
that a gravitational "charge" of a photon must be proportional to
$E$ (it gives the factor  $E^{2}$ in a cross-section) and a
normalization of a photon wave function gives the factor $E^{-1}$
in the cross-section. Also we assume here that a photon average
energy loss $\bar \epsilon $ in one act of interaction is
relatively small to a photon energy $E.$ Then average energy
losses of a photon with an energy  $E $ on a way $dr $ will be
equal to \cite{5,5a}:
\begin{equation}
                  dE=-aE dr,
\end{equation}
where $a$ is a constant. If a {\it whole} redshift magnitude is
caused by this effect, we must identify $a=H/c,$ where $c$ is the
light velocity, to have the Hubble law for small distances
\cite{104}.
\par
A photon energy $E$ should depend on a distance from a source $r$
as
\begin{equation}
                      E(r)=E_{0} \exp(-ar),
\end{equation}
where $E_{0}$ is an initial value of energy.
\par
The expression (2) is just only so far as the condition $\bar
\epsilon << E(r)$ takes place. Photons with a very small energy
may lose or acquire an energy changing their direction of
propagation after scattering.  Early or late such photons should
turn out in the thermodynamic equilibrium with the graviton
background, flowing into their own background. Decay of virtual
gravitons should give photon pairs for this background, too.
Perhaps, the last one is the cosmic microwave background \cite
{133,134}.
\par
It follows from the expression (2) that an exact dependence $r(z)$
is the following one:
\begin{equation}
                     r(z)= ln (1+z)/a,
\end{equation}
if an interaction with the graviton background is the only cause
of redshifts. It is very important, that this redshift does not
depend on a light frequency. For small $z,$ the dependence $r(z)$
will be linear.
\par
The expressions (1) - (3) are the same that appear in other
tired-light models (compare with \cite {212}). In this approach,
the ones follow from a possible existence of the isotropic
graviton background, from quantum electrodynamics, and from the
fact that a gravitational "charge" of a photon must be
proportional to $E.$
\subsection[2.2]{Non-forehead collisions with gravitons: an
additional dimming of any light flux} Photon flux's average energy
losses on a way $dr$ due to non-forehead collisions with gravitons
should be proportional to $badr,$ where $b$ is a new constant of
the order $1.$ These losses are connected with a rejection of a
part of photons from a source-observer direction. Such the
relaxation together with the redshift will give a connection
between visible object's diameter and its luminosity (i.e. the
ratio of an  object visible angular diameter to a square root of
visible luminosity), distinguishing from the one of the standard
cosmological model.
\par
Let us consider that in a case of a non-forehead collision of a
graviton with a photon, the latter leaves a photon flux detected
by a remote observer (an assumption of a narrow beam of rays). The
details of calculation of the theoretical value of relaxation
factor $b$ which was used in author's paper \cite{5} were given
later in the preprint \cite{213}. So as both particles have
velocities $c,$ a cross-section of interaction, which is "visible"
under an angle $\theta$ (see Fig. 1), will be equal to $\sigma_{0}
\vert \cos \theta \vert$ if $\sigma_{0}$ is a cross-section by
forehead collisions. The function $\vert \cos \theta \vert$ allows
to take into account both front and back hemispheres for riding
gravitons. Additionally, a graviton flux, which falls on a picked
out area (cross-section), depends on the angle $\theta.$ We have
for the ratio of fluxes:
$$\Phi(\theta)/\Phi_{0}=S_{s}/\sigma_{0}, $$
where $\Phi(\theta)$ and $\Phi_{0}$ are the fluxes which fall on
$\sigma_{0}$ under the angle $\theta$ and normally, $S_{s}$ is a
square of side surface of a truncated cone with a base
$\sigma_{0}$ (see Fig. 1).

\begin{figure}[th]
\epsfxsize=12.98cm \centerline{\epsfbox{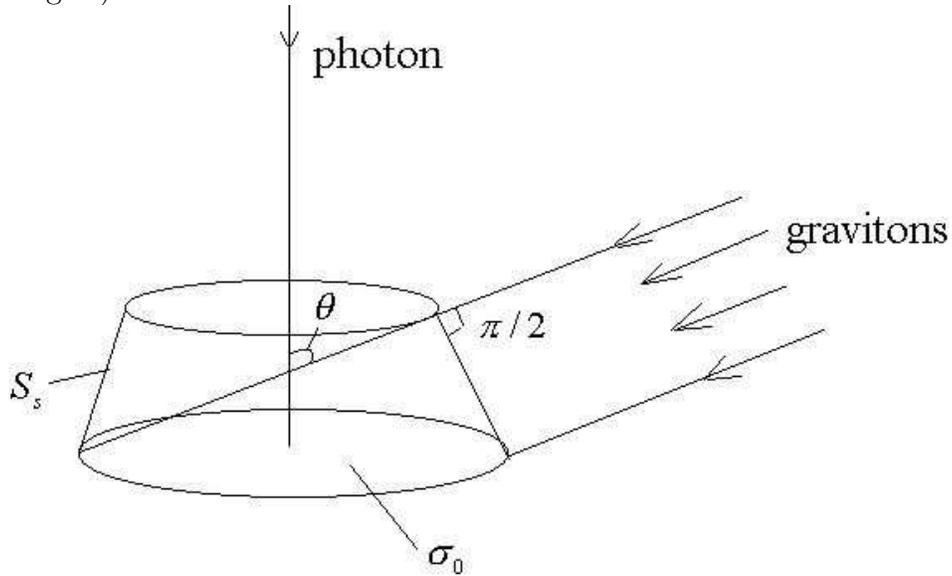}} \caption{By
non-forehead collisions of gravitons with a photon, it is
necessary to calculate a cone's side surface square, $S_{s}.$}
\end{figure}
Finally, we get for the factor $b:$
\begin{equation}
b=2 \int_{0}^{\pi/2}\cos{\theta}\cdot (S_{s}/\sigma_{0})\frac
{d\theta}{\pi/2}.
\end{equation}
By $0<\theta<\pi/4,$ a formed cone contains self-intersections,
and it is $S_{s}=2\sigma_{0} \cdot \cos{\theta}$. By
$\pi/4\leq\theta\leq\pi/2,$ we have $S_{s}=4\sigma_{0} \cdot
\sin^{2}{\theta}\cos{\theta}$.
\par
After computation of simple integrals, we get:
\begin{equation}
b=\frac {4}{\pi} (\int_{0}^{\pi/4}2\cos^{2}{\theta}d\theta +
\int_{\pi/4}^{\pi/2}\sin^{2}{2\theta}d\theta)= \frac {3}{2} +
\frac {2}{\pi} \simeq 2.137.
\end{equation}
In the considered simplest case of the uniform non-expanding
universe with the Euclidean space, we shall have the quantity
$$(1+z)^{(1+b)/2} \equiv (1+z)^{1.57}$$
in a visible object diameter-luminosity connection if a whole
redshift magnitude would caused by such an interaction with the
background (instead of $(1+z)^{2}$ for the expanding uniform
universe). For near sources, the estimate of the factor $b$ will
be some increased one.
\par
The luminosity distance (see \cite{2}) is a convenient quantity
for astrophysical observations. Both redshifts and the additional
relaxation of any photonic flux due to non-forehead collisions of
gravitons with photons lead in our model to the following
luminosity distance $D_{L}:$
\begin{equation}
D_{L}=a^{-1} \ln(1+z)\cdot (1+z)^{(1+b)/2} \equiv a^{-1}f_{1}(z),
\end{equation}
where $f_{1}(z)\equiv \ln(1+z)\cdot (1+z)^{(1+b)/2}$.
\subsection[2.3]{Comparison of the theoretical predictions with
supernova data }
 To compare a form of this predicted dependence
$D_{L}(z)$ by unknown, but constant $H$, with the latest
observational supernova data by Riess et al. \cite{203}, one can
introduce distance moduli $\mu_{0} = 5 \log D_{L} + 25 = 5 \log
f_{1} + c_{1}$, where $c_{1}$ is an unknown constant (it is a
single free parameter to fit the data); $f_{1}$ is the luminosity
distance in units of $c/H$.
\begin{figure}[th]
\epsfxsize=12.98cm \centerline{\epsfbox{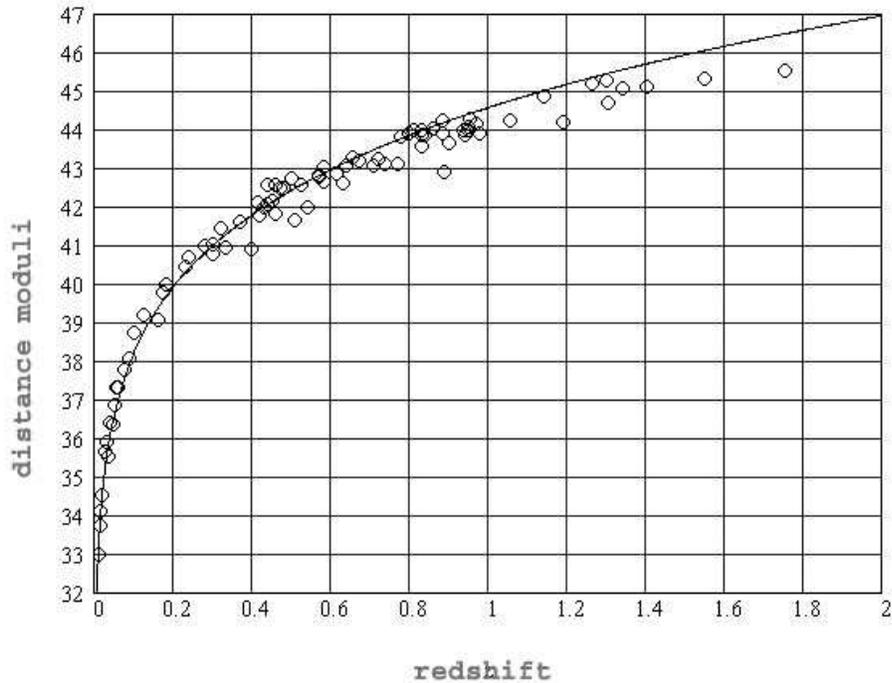}}
\caption{Comparison of the theoretical values of distance moduli
$\mu_{0}(z)$ (solid line) with observations (points) from
\cite{203} by Riess et al.}
\end{figure}
In Figure 2, the Hubble diagram $\mu_{0}(z)$ is shown with
$c_{1}=43$ to fit observations for low redshifts; observational
data (82 points) are taken from Table 5 of \cite{203}. The
predictions fit observations very well for roughly $z < 0.5$. It
excludes a need of any dark energy to explain supernovae dimming.
\par
Discrepancies between predicted and observed values of
$\mu_{0}(z)$ are obvious for higher $z$: we see that observations
show brighter SNe that the theory allows, and a difference
increases with $z$. It is better seen on Figure 3 with a linear
scale for $f_{1}$; observations are transformed as $\mu_{0}
\rightarrow 10^{(\mu_{0}-c_{1})/5}$ with the same
$c_{1}=43$.\footnote{A spread of observations raises with $z$; it
might be partially caused by quickly raising contribution of a
dispersion of measured flux: it should be proportional to
$f_{1}^{6}(z)$.}
\begin{figure}[th]
\epsfxsize=12.98cm \centerline{\epsfbox{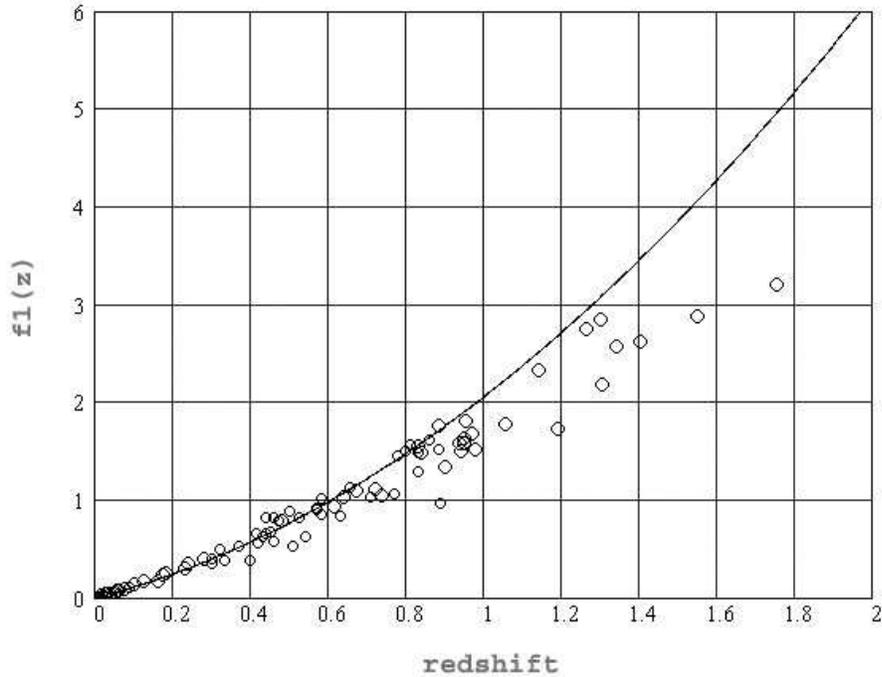}}
\caption{Predicted values of $f_{1}(z)$ (solid line) and
observations (points) from \cite{203} transformed to a linear
scale}
\end{figure}
\par
It would be explained in the model as a result of specific
deformation of SN spectra due to a discrete character of photon
energy losses. Today, a theory of this effect does not exist, and
I explain its origin only qualitatively \cite{111}. For very small
redshifts $z,$ only a small part of photons transmits its energy
to the background (see below Fig. 8 in Section 6). Therefore any
red-shifted narrow spectral strip will be a superposition of two
strips. One of them has a form which is identical with an initial
one, its space is proportional to $1-n(r)$ where $n(r)$ is an
average number of interactions of a single photon with the
background, and its center's shift is negligible (for a narrow
strip). Another part is expand, its space is proportional to
$n(r),$ and its center's shift is equal to $\bar \epsilon _{g} /h$
where $\bar \epsilon _{g} $   is an average energy loss in one act
of interaction. An amplitude of the red-shifted step should linear
raise with a redshift. For big $z,$ spectra of remote objects of
the universe would be deformed. A deformation would appear because
of multifold interactions of a initially-red-shifted part of
photons with the graviton background. It means that the observed
flux within a given passband would depend on a form of spectrum:
the flux may be larger than an expected one without this effect if
an initial flux within a next-blue neighbour band is big enough -
due to a superposition of red-shifted parts of spectrum. Some
other evidences of this effect would be an apparent variance of
the fine structure constant \cite{205} or of the CMB temperature
\cite{206} with epochs. In both cases, a ratio of red-shifted
spectral line's intensities may be sensitive to the effect.
\par
This comparison with observations is very important; to see some
additional details, we can compute and graph the ratio $f_{1
obs}(z)/f_{1}(z)$, where $f_{1 obs}(z)$ is an observed analog of
$f_{1}(z)$ (see Fig. 4) \cite{204}. An expected value of the ratio
should be equal to 1 for any $z$.
\begin{figure}[th]
\centerline{\includegraphics[width=12.98cm]{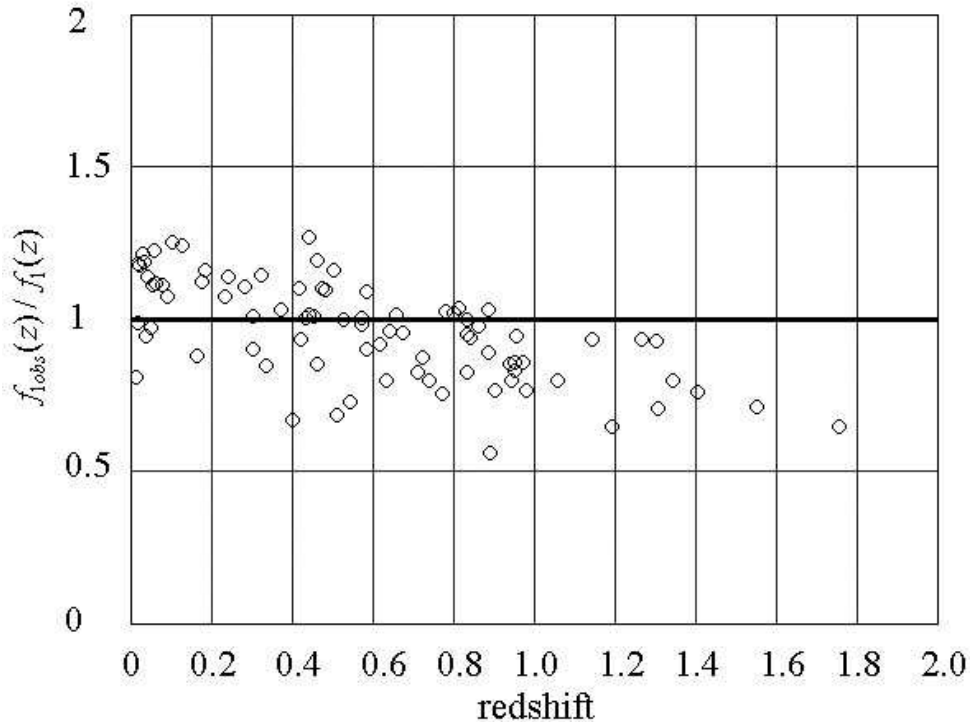}}
\caption{The ratio of observed to theoretical functions $f_{1
obs}(z)/f_{1}(z)$ (dots); observational data are taken from Table
5 of \cite{203}. If this model is true, the ratio should be equal
to 1 for any $z$ (solid line).}
\end{figure}

\subsection[2.4]{Computation of the Hubble constant}
Let us consider that a full redshift magnitude is caused by an
interaction with single gravitons. If $\sigma (E,\epsilon)$ is a
cross-section of interaction by forehead collisions of a photon
with an energy $E$ with a graviton, having an energy $\epsilon,$
we consider really (see (1)), that
$${d \sigma (E,\epsilon) \over E d \Omega} = const(E),$$
where $d \Omega$ is a space angle element, and the function
$const(x)$ has a constant value for any $x$. If $f(\omega,T)d
\Omega /2 \pi$ is a spectral density of graviton flux in the
limits of $d \Omega$ in some direction ($\omega$ is a graviton
frequency, $\epsilon =\hbar\omega$), i.e. an intensity of a
graviton flux is equal to the integral $ (d \Omega/2 \pi)
\int_{0}^{\infty} f(\omega,T)d \omega,$ $T$ is an equivalent
temperature of the graviton background, we can write for the
Hubble constant $H=ac,$ introduced in the expression (1):
$$H={1 \over 2\pi} \int_{0}^{\infty} \frac {\sigma (E,\epsilon)}{E}
f(\omega,T)d \omega.$$ If $f(\omega,T)$ can be described by the
Planck formula for equilibrium radiation, then $$
\int_{0}^{\infty} f(\omega,T)d \omega = \sigma T^{4},$$ where
$\sigma$ is the Stephan- Boltzmann constant. As carriers of a
gravitational "charge" (without consideration of spin properties),
gravitons should be described in the same manner as photons
(compare with \cite{419}), i.e. one can write for them:
$${d \sigma (E,\epsilon) \over \epsilon d\Omega} = const(\epsilon).$$
Now let us introduce a new dimensional constant $D$, so that for
forehead collisions:
\begin{equation}
\sigma (E,\epsilon)= D \cdot E \cdot \epsilon.
\end{equation}
Then
\begin{equation}
H= {1 \over 2\pi} D \cdot \bar \epsilon \cdot (\sigma T^{4}),
\end{equation}
where $\bar \epsilon$ is an average graviton energy. Assuming $T
\sim 3 K,~ \bar \epsilon \sim 10^{-4}~ eV,$ and $H = 1.6 \cdot
10^{-18}~ s^{-1},$ we get the following rough estimate for $D:$
$$D \sim 10^{-27}~ m^{2}/eV^{2},$$ (see below Section 4.3 for more
exact estimate of $D$ and for a theoretical estimate of $H$) that
gives us the phenomenological estimate of cross-section by the
same and equal $E$ and $\bar \epsilon$:
$$\sigma (E,\bar \epsilon) \sim 10^{-35}~ m^{2}.$$

\subsection[2.5]{Some new constants from dimensional reasonings}
We can introduce the following new constants (see \cite{5}):
$G_{0},~ l_{0},~ E_{0},$ which are analogues, on this new scale,
of classical constants: the Newton constant $G,$ the Planck length
$l_{Pl},$ and the Planck energy $E_{Pl}$ correspondingly. Let us
accept from dimensional reasonings that
$$D \equiv (l_{0}/ E_{0})^{2} = (G_{0}/c^{4})^{2},$$ where $
l_{0}=\sqrt{ G_{0} \hbar /c^{3}},~ E_{0}=\sqrt{ \hbar c^{5}/
G_{0}}.$ Then we have for these new constants: $$G_{0} \sim 1.6
\cdot 10^{39} m^{3}/kg \cdot s^{2},~ l_{0} \sim 2.4\cdot 10^{-12}
m,~ E_{0} \sim 1.6\ keV.$$ If one would replace $G$ with $G_{0},$
then an electrostatic force, acting between two protons, will be
only $\sim 2 \cdot 10^{13}$ times smaller than a gravitational one
by the same distance (a theoretical finding of the Newton constant
$G$ is given below in Section 4.3).
\par
Using $E_{0}$ instead of $E_{Pl},$ we can evaluate the new
non-dimensional "constant" (a bilinear function of $E$ and
$\epsilon$), $k,$ which would characterize one act of interaction:
$k \equiv E \cdot \epsilon / E_{0}^{2}. $ We must remember here,
that a universality of gravitational interaction allows to expect
that this floating coupling "constant" $k$ should characterize
interactions of any particles with an energy $E,$ including
gravitons, with single gravitons. For $E \sim 1 eV$ and $\epsilon
\sim 10^{-4} eV,$ we have $k  \sim 4 \cdot 10^{-9}.$ But for  $E
\sim 25~ MeV$ and  $\epsilon \sim 10^{-3}~ eV,$ we shall have $k
\sim 10^{-2},$ i.e. $k$  will be comparable with QED's constant
$\alpha.$ Already by $E \sim \epsilon \sim 5~ keV,$ such an
interaction would have the same intensity as the strong
interaction ($k \sim 10$).

\section[3]{Deceleration of massive bodies: an analog of redshifts}

As it was reported by Anderson's team \cite{1} , NASA deep-space
probes (Pioneer 10/11, Galileo, and Ulysses) experience a small
additional constant acceleration, directed towards the Sun (the
Pioneer anomaly). Today, a possible origin of the effect is
unknown. It must be noted here that the reported direction of
additional acceleration may be a result of the simplest
conjecture, which was accepted by the authors to provide a good
fit for all probes. One should compare different conjectures to
choose the one giving the best fit.
\par
We consider here a deceleration of massive bodies, which would
give a similar deformation of cosmic probes' trajectories
\cite{5}. The one would be a result of interaction of a massive
body with the graviton background, but such an additional
acceleration will be directed against a body velocity.
\par
It follows from a universality of gravitational interaction, that
not only photons, but all other objects, moving relative to the
background, should lose their energy, too, due to such a quantum
interaction with gravitons. If $a=H/c,$ it turns out that massive
bodies must feel a constant  deceleration of the same order of
magnitude as a small additional acceleration of cosmic probes.
\par
Let us now denote as $E$ a full energy of a moving body which has
a velocity $v$ relative to the background. Then energy losses of
the body by an interaction with the graviton background (due to
forehead collisions with gravitons) on the way $dr$ must be
expressed by the same formula (1):
$$ dE=-aE dr,$$
where $a=H/c.$ If $dr=vdt,$ where $t$ is a time, and
$E=mc^{2}/\sqrt{1-v^{2}/c^{2}},$ then we get for the body
acceleration $w \equiv dv/dt$ by a non-zero velocity:
\begin{equation}
w = - ac^{2}(1-v^{2}/c^{2}).
\end{equation}
We assume here, that non-forehead collisions with gravitons give
only stochastic deviations of a  massive body's velocity
direction, which are negligible. For small velocities:
\begin{equation}
w \simeq - Hc.
\end{equation}
If the Hubble constant $H$ is equal to $2.14 \cdot 10^{-18}
s^{-1}$ (it is the theoretical estimate of $H$ in this approach,
see below Section 4.3), a modulus of the acceleration will be
equal to
\begin{equation}
|w| \simeq Hc = 6.419 \cdot 10^{-10} \ m/s^{2},
\end{equation}
that has the same order of magnitude as  a value of the observed
additional acceleration $(8.74 \pm 1.33) \cdot 10^{-10} m/s^2$ for
NASA probes.
\par
We must emphasize here that the acceleration $w$ is directed
against a body velocity only in a special frame of reference (in
which the graviton background is isotropic). In other frames, we
may find its direction, using transformation formulae for an
acceleration (see \cite{309}). We can assume that the graviton
background and the microwave one are isotropic in one frame (the
Earth velocity relative to the microwave background was determined
in \cite{316}).
\par
To verify my conjecture about the origin of probes' additional
acceleration, one could re-analyze radio Doppler data for the
probes. One should find a velocity of the special frame of
reference and a constant probe acceleration $w$ in this frame
which must be negative, as it is described above. These two
parameters must provide the best fit for all probes, if the
conjecture is true. In such a case, one can get an independent
estimate of the Hubble constant, based on the measured value of
probe's additional acceleration: $H= \mid w \mid /c.$ I would like
to note that a deep-space mission to test the discovered anomaly
is planned now at NASA by the authors of this very important
discovery \cite{317}.
\par
Under influence of such a small additional acceleration $w$, a
probe must move on a deformed trajectory.  Its view will be
determined by small seeming deviations from exact conservation
laws for energy and angular momentum of a not-fully reserved body
system which one has in a case of neglecting with the graviton
background. For example, Ulysses should go some nearer to the Sun
when the one rounds it. It may be interpreted as an additional
acceleration, directed towards the Sun, if we shall think that one
deals with a reserved body system.
\par
It is very important to understand, why such an acceleration has
not been observed for planets. This acceleration will have
different directions by motion of a body on a closed orbit, and
one must take into account a solar system motion, too. As a
result, an orbit should be deformed. The observed value of
anomalous acceleration of Pioneer 10 should represent the vector
difference of the two accelerations \cite{6}: an acceleration of
Pioneer 10 relative to the graviton background, and  an
acceleration of the Earth relative to the background. Possibly,
the last is displayed as an annual periodic term in the residuals
of Pioneer 10 \cite{311}. If the solar system moves with a
noticeable velocity relative to the background, the Earth's
anomalous acceleration projection on the direction of this
velocity will be smaller than for the Sun - because of the Earth's
orbital motion. It means that in a frame of reference, connected
with the Sun, the Earth should move with an anomalous acceleration
having non-zero projections as well on the orbital velocity
direction as on the direction of solar system motion relative to
the background. Under some conditions, the Earth's anomalous
acceleration in this frame of reference may be periodic. The axis
of Earth's orbit should feel an annual precession by it. This
question needs a further consideration.
\section[4]{Gravity as the screening effect}
It was shown by the author \cite{6,06,006} that screening the
background of super-strong interacting gravitons creates for any
pair of bodies both attraction and repulsion forces due to
pressure of gravitons. For single gravitons, these forces are
approximately balanced, but each of them is much bigger than a
force of Newtonian attraction. If single gravitons are pairing, an
attraction force due to pressure of such graviton pairs is twice
exceeding a corresponding repulsion force if graviton pairs are
destructed by collisions with a body. In such the model, the
Newton constant is connected with the Hubble constant that gives a
possibility to obtain a theoretical estimate of the last. We deal
here with a flat non-expanding universe fulfilled with
super-strong interacting gravitons; it changes the meaning of the
Hubble constant which describes magnitudes of three small effects
of quantum gravity but not any expansion or an age of the
universe.
\subsection[4.1]{Pressure force of single gravitons}
If gravitons of the background run against a pair of bodies with
masses $m_{1}$ and $m_{2}$ (and energies $E_{1}$ and $E_{2}$) from
infinity, then a part of gravitons is screened. Let $\sigma
(E_{1},\epsilon)$ is a cross-section of interaction of body $1$
with a graviton with an energy $\epsilon=\hbar \omega,$ where
$\omega$ is a graviton frequency, $\sigma (E_{2},\epsilon)$ is the
same cross-section for body $2.$ In absence of body $2,$ a whole
modulus of a gravitonic pressure force acting on body $1$ would be
equal to:
\begin{equation}
4\sigma (E_{1},<\epsilon>)\cdot {1 \over 3} \cdot {4 f(\omega, T)
\over c},
\end{equation}
where $f(\omega, T)$ is a graviton spectrum with a temperature $T$
(assuming to be Planckian), the factor $4$ in front of $\sigma
(E_{1},<\epsilon>)$ is introduced to allow all possible directions
of graviton running, $<\epsilon>$ is another average energy of
running gravitons with a frequency $\omega$ taking into account a
probability of that in a realization of flat wave a number of
gravitons may be equal to zero, and that not all of gravitons ride
at a body.
\par
Body $2,$ placed on a distance $r$ from body $1,$ will screen a
portion of running against body $1$ gravitons which is equal for
big distances between the bodies (i.e. by $\sigma
(E_{2},<\epsilon>) \ll 4 \pi r^{2}$) to:
\begin{equation}
\sigma (E_{2},<\epsilon>) \over 4 \pi r^{2}.
\end{equation}
Taking into account all frequencies $\omega,$ the following
attractive force will act between bodies $1$ and $2:$
\begin{equation}
F_{1}= \int_{0}^{\infty} {\sigma (E_{2},<\epsilon>) \over 4 \pi
r^{2}} \cdot 4 \sigma (E_{1},<\epsilon>)\cdot {1 \over 3} \cdot {4
f(\omega, T) \over c} d\omega.
\end{equation}
Let $f(\omega, T)$ is described with the Planck formula:
\begin{equation}
f(\omega,T)={{\omega}^{2} \over {4{\pi}^{2} c^{2}}} {{\hbar
\omega} \over {\exp(\hbar \omega/kT) - 1}}.
\end{equation}
Let $x \equiv {\hbar \omega/  kT},$ and $\bar{n} \equiv {1/
(\exp(x)-1)}$ is an average number of gravitons in a flat wave
with a frequency $\omega$ (on one mode of two distinguishing with
a projection of particle spin). Let $P(n,x)$ is a probability of
that in a realization of flat wave a number of gravitons is equal
to $n,$ for example $P(0,x)=\exp(-\bar{n}).$
\par
A quantity $<\epsilon>$ must contain the factor $(1-P(0,x)),$ i.e.
it should be:
\begin{equation}
<\epsilon> \sim \hbar \omega (1-P(0,x)),
\end{equation}
that let us to reject flat wave realizations with zero number of
gravitons.
\par
But attempting to define other factors in $<\epsilon>,$ we find
the difficult place in our reasoning. On this stage, it is
necessary to introduce some new assumption to find the factors.
Perhaps, this assumption will be well-founded in a future theory -
or would be rejected. If a flat wave realization, running against
a finite size body from infinity, contains one graviton, then one
cannot consider that it must stringent ride at a body to interact
with some probability with the one. It would break the uncertainty
principle by W. Heisenberg. We should admit that we know a
graviton trajectory. The same is pertaining to gravitons scattered
by one of bodies by big distances between bodies. What is a
probability that a single graviton will ride namely at the body?
If one denotes this probability as $P_{1},$ then for a wave with
$n$ gravitons their chances to ride at the body must be equal to
$n \cdot P_{1}.$ Taking into account the probabilities of values
of $n$ for the Poisson flux of events, an additional factor in
$<\epsilon>$ should be equal to $\bar{n} \cdot P_{1}.$ I have
admitted in \cite{6} that
\begin{equation}
P_{1}=P(1,x),
\end{equation}
where $P(1,x)=\bar{n}\exp(-\bar{n});$ (below it is admitted for
pairing gravitons: $P_{1}=P(1,2x)$ - see Section 4.3).
\par
In such the case, we have for $<\epsilon>$ the following
expression:
\begin{equation}
<\epsilon>= \hbar \omega (1-P(0,x))\bar{n}^{2}\exp(-\bar{n}).
\end{equation}
Then we get for an attractive force $F_{1}:$
\begin{equation}
F_{1}= {4 \over 3}  {{D^{2} E_{1} E_{2}} \over {\pi r^{2} c}}
\int_{0}^{\infty} {{{\hbar}^{3} \omega^{5}} \over {4\pi^{2}c^{2}}}
(1-P(0,x))^{2}\bar{n}^{5}\exp(-2\bar{n}) d\omega =
\end{equation}
$${1 \over 3} \cdot {{D^{2} c (kT)^{6} m_{1} m_{2}} \over
{\pi^{3}\hbar^{3}r^{2}}} \cdot I_{1},$$ where
\begin{equation}
I_{1} \equiv \int_{0}^{\infty} x^{5}
(1-\exp(-(\exp(x)-1)^{-1}))^{2}(\exp(x)-1)^{-5}
\exp(-2(\exp(x)-1)^{-1}) dx=
\end{equation}
$$5.636 \cdot 10^{-3}.$$ This and all other integrals were found
with the MathCad software.

If $F_{1}\equiv G_{1} \cdot  m_{1}m_{2}/r^{2},$ then the constant
$G_{1}$ is equal to:
\begin{equation}
G_{1} \equiv {1 \over 3} \cdot {D^{2} c(kT)^{6} \over
{\pi^{3}\hbar^{3}}} \cdot I_{1}.
\end{equation}
By $T=2.7~ K:$
\begin{equation}
G_{1} =1215.4 \cdot G,
\end{equation}
that is three order greater than the Newton constant, $G.$
\par
But if single gravitons are elastically scattered with body $1,$
then our reasoning may be reversed: the same portion (13) of
scattered gravitons will create a repulsive force $F_{1}^{'}$
acting on body $2$ and equal to
\begin{equation}
F_{1}^{'} =F_{1},
\end{equation}
if one neglects with small allowances which are proportional to
$D^{3}/  r^{4}.$
\par
So, for bodies which elastically scatter gravitons, screening a
flux of single gravitons does not ensure Newtonian attraction. But
for gravitonic black holes which absorb any particles and do not
re-emit them (by the meaning of a concept, the ones are usual
black holes; I introduce a redundant adjective only from a
caution), we will have $F_{1}^{'} =0.$ It means that such the
object would attract other bodies with a force which is
proportional to $G_{1}$ but not to $G,$ i.e. Einstein's
equivalence principle would be violated for them. This conclusion,
as we shall see below, stays in force for the case of graviton
pairing, too. The conclusion cannot be changed with taking into
account of Hawking's  quantum effect of evaporation of black holes
\cite{010}.
\subsection[4.2]{Graviton pairing}
To ensure an attractive force which is not equal to a repulsive
one, particle correlations should differ for {\it in} and {\it
out} flux. For example, single gravitons of running flux may
associate in pairs \cite{6}. If such pairs are destructed by
collision with a body, then quantities $<\epsilon>$ will be
distinguished for running and scattered particles. Graviton
pairing may be caused with graviton's own gravitational attraction
or gravitonic spin-spin interaction. Left an analysis of the
nature of graviton pairing for the future; let us see that gives
such the pairing.
\par
To find an average number of pairs $\bar{n}_{2}$ in a wave with a
frequency $\omega$ for the state of thermodynamic equilibrium, one
may replace $\hbar \rightarrow 2\hbar$ by deducing the Planck
formula. Then an average number of pairs will be equal to:
\begin{equation}
\bar{n}_{2} ={1 \over {\exp(2x)-1}},
\end{equation}
and an energy of one pair will be equal to $2\hbar \omega.$ It is
important that graviton pairing does not change a number of
stationary waves, so as pairs nucleate from existing gravitons.
The question arises: how many different modes, i.e. spin
projections, may graviton pairs have? We consider that the
background of initial gravitons consists of two modes. For
massless transverse bosons, it takes place as by spin $1$ as by
spin $2.$ If graviton pairs have maximum spin $2,$ then single
gravitons should have spin $1.$ But from such particles one may
constitute four combinations: $\uparrow \uparrow, \ \downarrow
\downarrow $ (with total spin $2$), and $\uparrow \downarrow, \
\downarrow\uparrow$ (with total spin $0).$ All these four
combinations will be equiprobable if spin projections $\uparrow$
and $\downarrow$ are equiprobable in a flat wave (without taking
into account a probable spin-spin interaction).
\par
But it is happened that, if expression (24)  is true, it follows
from the energy conservation law that composite gravitons should
be distributed only in two modes. So as
\begin{equation}
\lim_{x \to 0} {\bar{n}_{2} \over \bar{n}} ={1/2},
\end{equation}
then by $x \rightarrow 0$ we have $2\bar{n}_{2}=\bar{n},$ i.e. all
of gravitons are pairing by low frequencies. An average energy on
every mode of pairing gravitons is equal to $2 \hbar \omega
\bar{n}_{2},$ the one on every mode of single gravitons - to
$\hbar \omega \bar{n}.$ These energies are equal by $x \rightarrow
0,$ because of that, the numbers of modes are equal, too, if the
background is in the thermodynamic equilibrium with surrounding
bodies.
\par
The above reasoning does not allow to choose a spin value $2$ or
$0$ for composite gravitons. A choice of namely spin $2$ would
ensure the following proposition: all of gravitons in one
realization of flat wave have the same spin projections. From
another side, a spin-spin interaction would cause it.
\par
The spectrum of composite gravitons is also the Planckian one, but
with a smaller temperature; it has the view:
\begin{equation}
f_{2}(2\omega,T)d\omega={\omega^{2} \over {4\pi^{2}c^{2}}} \cdot
{2\hbar \omega \over {\exp(2x)-1}}d\omega \equiv {(2\omega)^{2}
\over {32\pi^{2}c^{2}}} \cdot {2\hbar \omega \over
{\exp(2x)-1}}d(2\omega).
\end{equation}
It means that an absolute luminosity for the sub-system of
composite gravitons is equal to:
\begin{equation}
\int_{0}^{\infty} f_{2}(2\omega,T)d(2\omega)= {1 \over 8}\sigma
T^{4},
\end{equation}
where $\sigma$ is the Stephan-Boltzmann constant; i.e. an
equivalent temperature of this sub-system is
\begin{equation}
T_{2} \equiv (1/8)^{1/4}T = {2^{1/4} \over 2}T = 0.5946T.
\end{equation}
The portion of pairing gravitons, $2\bar n_{2}/\bar n,$  a
spectrum of single gravitons, $f(x),$ and a spectrum of subsystem
of pairing gravitons, $f_{2}(2x),$ are shown on Fig. 5 as
functions of the dimensionless parameter $x \equiv \hbar \omega
/kT$.
\par
\begin{figure}[th]
\centerline{\includegraphics[width=12.98cm]{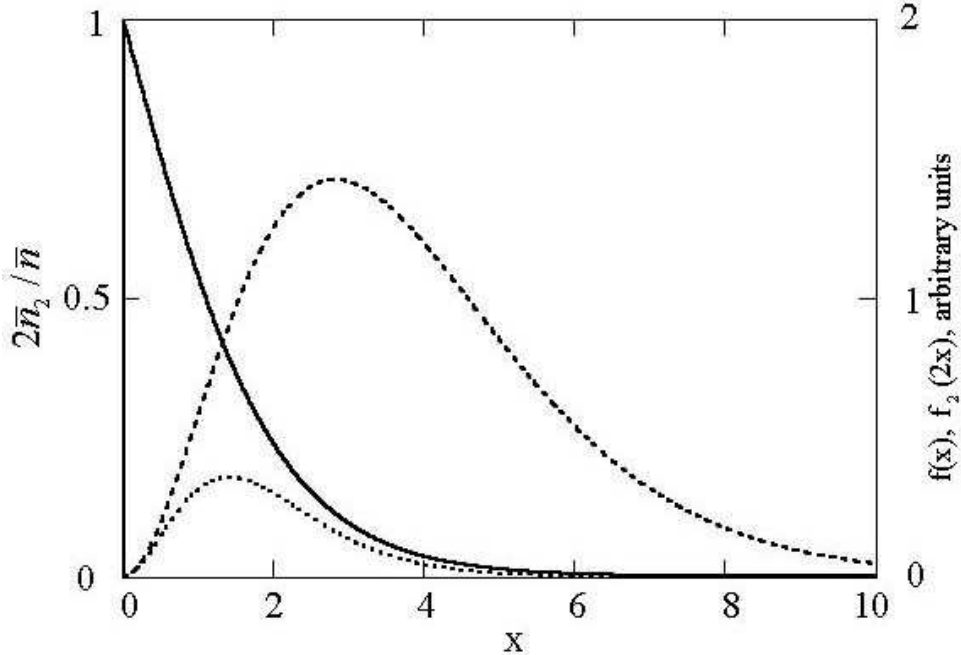}}
\caption{The portion of pairing gravitons, $2\bar n_{2}/\bar n,$
(solid line), a spectrum of single gravitons, $f(x),$ (dashed
line), and a spectrum of graviton pairs, $f_{2}(2x),$ (dotted
line) versus the dimensionless parameter $x$.}
\end{figure}

It is important that the graviton pairing effect does not change
computed values of the Hubble constant and of anomalous
deceleration of massive bodies: twice decreasing of a sub-system
particle number due to the pairing effect is compensated with
twice increasing the cross-section of interaction of a photon or
any body with such the composite gravitons. Non-pairing gravitons
with spin $1$ give also its contribution in values of redshifts,
an additional relaxation of light intensity due to non-forehead
collisions with gravitons, and  anomalous deceleration of massive
bodies moving relative to the background.

\subsection[4.3]{Computation of the Newton constant, and a connection
between the two fundamental constants, $G$ and $H$ } If running
graviton pairs ensure for two bodies an attractive force $F_{2},$
then a repulsive force due to re-emission of gravitons of a pair
alone will be equal to $F_{2}^{'} =F_{2}/2.$ It follows from that
the cross-section for {\it single additional scattered} gravitons
of destructed pairs will be twice smaller than for pairs
themselves (the leading factor $2\hbar \omega$ for pairs should be
replaced with $\hbar \omega$ for single gravitons). For pairs, we
introduce here the cross-section $ \sigma (E_{2},<\epsilon_{2}>),$
where $<\epsilon_{2}>$ is an average pair energy with taking into
account a probability of that in a realization of flat wave a
number of graviton pairs may be equal to zero, and that not all of
graviton pairs ride at a body ($<\epsilon_{2}>$ is an analog of
$<\epsilon>$). This equality is true in neglecting with small
allowances which are proportional to $D^{3}/ r^{4}$ (see Section
4.4). Replacing $\bar{n} \rightarrow \bar{n}_{2},~ \hbar \omega
\rightarrow 2\hbar \omega,$ and $P(n,x) \rightarrow P(n,2x),$
where $P(0,2x)= \exp(-\bar{n}_{2}),$ we get for graviton pairs:
\begin{equation}
<\epsilon_{2}> \sim 2\hbar \omega
(1-P(0,2x))\bar{n}_{2}^{2}\exp(-\bar{n}_{2}).
\end{equation}
This expression does not take into account only that beside pairs
there may be single gravitons in a realization of flat wave. To
reject cases when, instead of a pair, a single graviton runs
against a body (a contribution of such gravitons in attraction and
repulsion is the same), we add the factor $P(0,x)$ into
$<\epsilon_{2}>:$
\begin{equation}
<\epsilon_{2}> = 2\hbar \omega
(1-P(0,2x))\bar{n}_{2}^{2}\exp(-\bar{n}_{2}) \cdot P(0,x).
\end{equation}
Then a force of attraction of two bodies due to pressure of
graviton pairs, $F_{2}$, - in the full analogy with (19) - will be
equal to \footnote{In initial version of this paper, factor 2 was
lost in the right part of Eq. (31), and the theoretical values of
$D$ and $H$ were overestimated of $\sqrt{2}$ times}:
\begin{equation}
F_{2}= \int_{0}^{\infty} {\sigma (E_{2},<\epsilon_{2}>) \over 4
\pi r^{2}} \cdot 4 \sigma (E_{1},<\epsilon_{2}>)\cdot {1 \over 3}
\cdot {4 f_{2}(2\omega,T) \over c} d\omega =
\end{equation}
$$ {8 \over 3} \cdot
{D^{2} c(kT)^{6} m_{1}m_{2} \over {\pi^{3}\hbar^{3}r^{2}}}\cdot
I_{2},$$ where
\begin{equation}
I_{2} \equiv \int_{0}^{\infty}{ x^{5}
(1-\exp(-(\exp(2x)-1)^{-1}))^{2}(\exp(2x)-1)^{-5} \over
\exp(2(\exp(2x)-1)^{-1}) \exp(2(\exp(x)-1)^{-1})} d x =
\end{equation}
$$2.3184 \cdot 10^{-6}.$$
The difference $F$ between attractive and repulsive forces will be
equal to:
\begin{equation}
F \equiv F_{2}- F_{2}^{'}={1 \over 2}F_{2} \equiv G_{2}{m_{1}m_{2}
\over r^{2}},
\end{equation}
where the constant $G_{2}$ is equal to:
\begin{equation}
G_{2} \equiv {4 \over 3} \cdot {D^{2} c(kT)^{6} \over
{\pi^{3}\hbar^{3}}} \cdot I_{2}.
\end{equation}
Both $G_{1}$ and $G_{2}$ are proportional to $T^{6}$ (and $H \sim
T^{5},$ so as $\bar{\epsilon} \sim T$).
\par
If one assumes that $G_{2}=G,$ then it follows from (34) that by
$T=2.7K$ the constant $D$ should have the value:
\begin{equation}
D=0.795 \cdot 10^{-27}{m^{2} / eV^{2}}.
\end{equation}
An average graviton energy of the background is equal to:
\begin{equation}
\bar{\epsilon} \equiv \int_{0}^{\infty} \hbar \omega \cdot
{f(\omega, T) \over \sigma T^{4}} d \omega = {15 \over
\pi^{4}}I_{4}kT,
\end{equation}
where
$$I_{4} \equiv \int_{0}^{\infty} {x^{4} dx \over
{\exp(x)-1}}=24.866 $$ (it is $\bar{\epsilon}=8.98 \cdot
10^{-4}eV$ by $T=2.7K$).
\par
We can use (8) and (34) to establish a connection between the two
fundamental constants, $G$ and $H$, under the condition that
$G_{2}=G.$ We have for $D:$
\begin{equation}
D= {2\pi H \over \bar{\epsilon} \sigma T^{4}}= {2 \pi^{5} H \over
15 k \sigma T^{5} I_{4}};
\end{equation}
then
\begin{equation}
G=G_{2} = {4 \over 3} \cdot {D^{2} c(kT)^{6} \over
{\pi^{3}\hbar^{3}}} \cdot I_{2}= \\
{64 \pi^{5} \over 45} \cdot {H^{2}c^{3}I_{2} \over \sigma T^{4}
I_{4}^{2}}.
\end{equation}
So as the value of $G$ is known much better than the value of $H,$
let us express $H$ via $G:$
\begin{equation}
H= (G  {45 \over 64 \pi^{5}}  {\sigma T^{4} I_{4}^{2} \over
{c^{3}I_{2}}})^{1/2}= 2.14 \cdot 10^{-18}~s^{-1},
\end{equation}
or in the units which are more familiar for many of us: $H=66.875
\ km \cdot s^{-1} \cdot Mpc^{-1}.$
\par
This value of $H$ is in the good accordance with the majority of
present astrophysical estimations \cite{2,512,513} (for example,
the estimate $(72 \pm 8)$ km/s/Mpc has been got from SN1a
cosmological distance determinations in \cite{513}), but it is
lesser than some of them \cite{512a} and than it follows from the
observed value of anomalous acceleration of Pioneer 10 \cite{1}.

\subsection[4.4]{Restrictions on a geometrical language in gravity}
The described quantum mechanism of classical gravity gives
Newton's law with the constant $G_{2}$ value (34) and the
connection (38) for the constants $G_{2}$ and $H.$  We have
obtained the rational value of $H$ (39) by $G_{2} = G,$ if the
condition of big distances is fulfilled:
\begin{equation}
\sigma (E_{2},<\epsilon>) \ll 4 \pi r^{2}.
\end{equation}
Because it is known from experience that for big bodies of the
solar system, Newton's law is a very good approximation, one would
expect that the condition (40) is fulfilled, for example, for the
pair Sun-Earth. But assuming $r=1 \ AU$ and
$E_{2}=m_{\odot}c^{2},$ we obtain assuming for rough estimation
$<\epsilon> \rightarrow \bar{\epsilon}:$ $${\sigma
(E_{2},<\epsilon>) \over 4 \pi r^{2}} \sim 4 \cdot 10^{12}. $$ It
means that in the case of interaction of gravitons or graviton
pairs with the Sun in the aggregate, the considered quantum
mechanism of classical gravity could not lead to Newton's law as a
good approximation. This "contradiction" with experience is
eliminated if one assumes that gravitons interact with "small
particles" of matter - for example, with atoms. If the Sun
contains of $N$ atoms, then $\sigma (E_{2},<\epsilon>)=N \sigma
(E_{a},<\epsilon>),$ where $E_{a}$ is an average energy of one
atom. For rough estimation we assume here that $E_{a}=E_{p},$
where $E_{p}$ is a proton rest energy; then it is $N \sim
10^{57},$ i.e. ${\sigma (E_{a},<\epsilon>)/ 4 \pi r^{2}} \sim
10^{-45} \ll 1.$
\par
This necessity of "atomic structure" of matter for working the
described quantum mechanism is natural relative to usual bodies.
But would one expect that black holes have a similar structure? If
any radiation cannot be emitted with a black hole, a black hole
should interact with gravitons as an aggregated object, i.e. the
condition (40) for a black hole of sun mass has not been fulfilled
even at distances $\sim 10^{6} \ AU.$
\par
For bodies without an atomic structure, the allowances, which are
proportional to $D^{3}/ r^{4}$ and are caused by decreasing a
gravitonic flux due to the screening effect, will have a factor
$m_{1}^{2}m_{2}$ or $m_{1}m_{2}^{2}.$ These allowances break the
equivalence principle for such the bodies.
\par
For bodies with an atomic structure, a force of interaction is
added up from small forces of interaction of their "atoms": $$ F
\sim N_{1}N_{2}m_{a}^{2}/r^{2}=m_{1}m_{2}/r^{2},$$ where $N_{1}$
and $N_{2}$ are numbers of atoms for bodies $1$ and $2$. The
allowances to full forces due to the screening effect will be
proportional to the quantity: $N_{1}N_{2}m_{a}^{3}/r^{4},$ which
can be expressed via the full masses of bodies as
$m_{1}^{2}m_{2}/r^{4}N_{1}$ or $m_{1}m_{2}^{2}/r^{4}N_{2}.$ By big
numbers $N_{1}$ and $N_{2}$ the allowances will be small. The
allowance to the force $F,$ acting on body $2,$ will be equal to:
\begin{equation}
\Delta F ={1 \over 2 N_{2}} \int_{0}^{\infty} {\sigma^{2}
(E_{2},<\epsilon_{2}>) \over (4 \pi r^{2})^{2}} \cdot 4 \sigma
(E_{1},<\epsilon_{2}>)\cdot {1 \over 3} \cdot {4 f_{2}(2\omega,T)
\over c} d\omega =
\end{equation}
$${2 \over 3N_{2}} \cdot {{D^{3} c^{3} (kT)^{7}
m_{1} m_{2}^{2}} \over {\pi^{4}\hbar^{3}r^{4}}} \cdot I_{3},$$
(for body $1$ we shall have the similar expression if replace
$N_{2} \rightarrow N_{1}, \ m_{1}m_{2}^{2} \rightarrow
m_{1}^{2}m_{2}$), where
$$ I_{3} \equiv \int_{0}^{\infty} {x^{6}
(1-\exp(-(\exp(2x)-1)^{-1}))^{3}(\exp(2x)-1)^{-7} \over
\exp(3(\exp(x)-1)^{-1})} d x = 1.0988 \cdot 10^{-7}. $$
\par
Let us find the ratio:
\begin{equation}
{\Delta F \over F} = {D E_{2} kT \over {N_{2} 2\pi r^{2}}} \cdot
{I_{3} \over I_{2}}.
\end{equation}
Using this formula, we can find by $E_{2}=E_{\odot}, \ r=1 \ AU:$
\begin{equation}
{\Delta F \over F} \sim 10^{-46}.
\end{equation}
\par
An analogical allowance to the force $F_{1}$ has by the same
conditions the order $\sim 10^{-48}F_{1},$ or $\sim 10^{-45}F.$
One can replace $E_{p}$ with a rest energy of very big atom - the
geometrical approach will left a very good language to describe
the solar system. We see that for bodies with an atomic structure
the considered mechanism leads to very small deviations from
Einstein's equivalence principle, if the condition (40) is
fulfilled for microparticles, which prompt interact with
gravitons.
\par
For small distances we shall have:
\begin{equation}
\sigma (E_{2},<\epsilon>) \sim 4 \pi r^{2}.
\end{equation}
It takes place by $E_{a}=E_{p}, \ <\epsilon> \sim 10^{-3} \ eV$
for $r \sim 10^{-11} \ m.$ This quantity is many orders larger
than the Planck length. The equivalence principle should be broken
at such distances.
\par
Under the condition (44), big digressions from Newton's law will
be caused with two factors: 1) a screening portion of a running
flux of gravitons is not small and it should be taken into account
by computation of the repulsive force; 2) a value of this portion
cannot be defined by the expression (13).
\par
Instead of (13), one might describe this portion at small
distances with an expression of the kind:
\begin{equation}
{1 \over 2}(1+ {\sigma (E_{a},<\epsilon>)/ \pi r^{2}}-(1+ {\sigma
(E_{a},<\epsilon>)/ \pi r^{2}})^{1/2} )
\end{equation}
(the formula for a spheric segment area is used here \cite{013}).
Formally, by ${\sigma (E_{a},<\epsilon>)/ \pi r^{2}} \rightarrow
\infty$ we shall have for the portion (45):
$$\sim {1 \over 2}({\sigma (E_{a},<\epsilon>)/ \pi r^{2}}-({\sigma
(E_{a},<\epsilon>)/ \pi})^{1/2}/r),$$ where the second term shows
that the interaction should be weaker at small distances.  We
might expect that a screening portion may tend to a fixing value
at super-short distances, and it will be something similar to
asymptotic freedom of strong interactions. But, of course, at such
distances the interaction will be super-strong and our naive
approach would be not valid.

\section[5]{Some cosmological consequences of the model}
If the described model of redshifts is true, what is a picture of
the universe? It is interesting that in a frame of this model,
every observer has two own spheres of observability in the
universe (two different cosmological horizons exist for any
observer) \cite{44,55}. One of them is defined by maximum existing
temperatures of remote sources - by big enough distances, all of
them will be masked with the CMB radiation. Another, and much
smaller, sphere depends on their maximum luminosity - the
luminosity distance increases with a redshift much quickly than
the geometrical one. The ratio of the luminosity distance to the
geometrical one is the quickly increasing function of $z:$
\begin{equation}
D_{L}(z)/r(z)= (1+z)^{(1+b)/2},
\end{equation}
which does not depend on the Hubble constant.  An outer part of
the universe will drown in a darkness. \par By the found
theoretical value of the Hubble constant: $H= 2.14 \cdot
10^{-18}~s^{-1}$ (then a natural light unit of distances is equal
to $1/H \simeq 14.85$ light GYR), plots of two theoretical
functions of $z$ in this model - the geometrical distance $r(z)$
and the luminosity distance $D_{L}(z)$ - are shown on Fig. 6
\cite{44,55}. As one can see, for objects with $z \sim 10$, which
are observable now, we should anticipate geometrical distances of
the order $\sim 35$ light GYR and luminosity distances of the
order $\sim 1555$ light GYR in a frame of this model. An estimate
of distances to objects with given $z$ is changed, too: for
example, the quasar with $z=5.8$ \cite {043} should be in a
distance approximately of 2.8 times bigger than the one expected
in the model based on the Doppler effect.
\begin{figure}[th]
\epsfxsize=12.98cm \centerline{\epsfbox{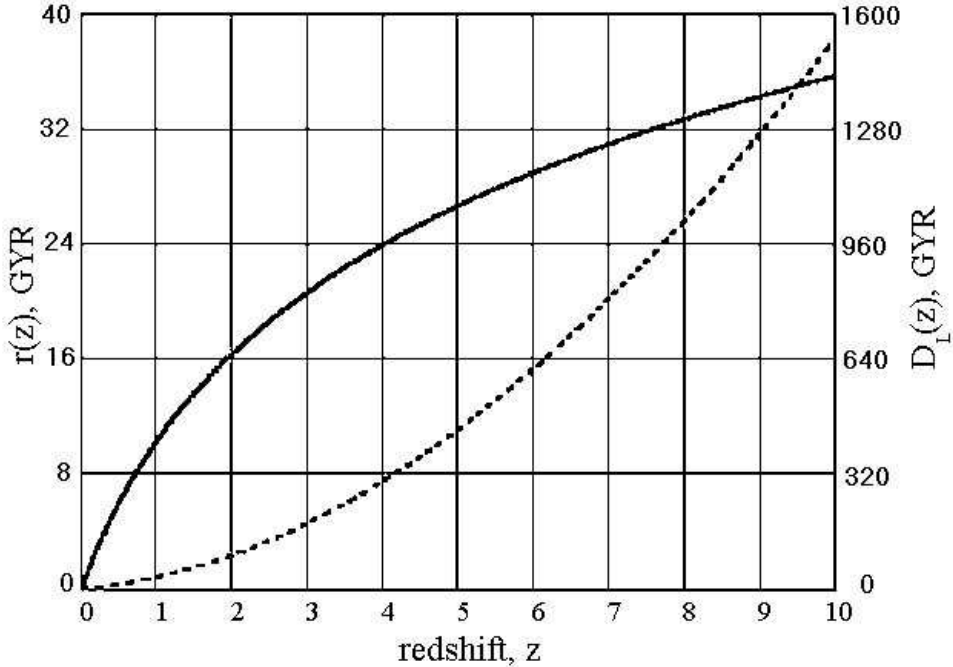}} \caption{The
geometrical distance, $r(z),$ (solid line) and the luminosity
distance, $D_{L}(z),$ (dashed line) - both in light GYRs - in this
model as functions of a redshift, z. The following theoretical
value for $H$ is accepted: $H= 2.14 \cdot 10^{-18}s^{-1}$.}
\end{figure}
\par
We can assume that the graviton background and the cosmic
microwave one are in a state of thermodynamical equilibrium, and
have the same temperatures. CMB itself may arise as a result of
cooling any light radiation up to reaching this equilibrium. Then
it needs $z \sim 1000$ to get through the very edge of our cosmic
"ecumene" (see Fig. 7).
\begin{figure}[th]
\epsfxsize=12.98cm \centerline{\epsfbox{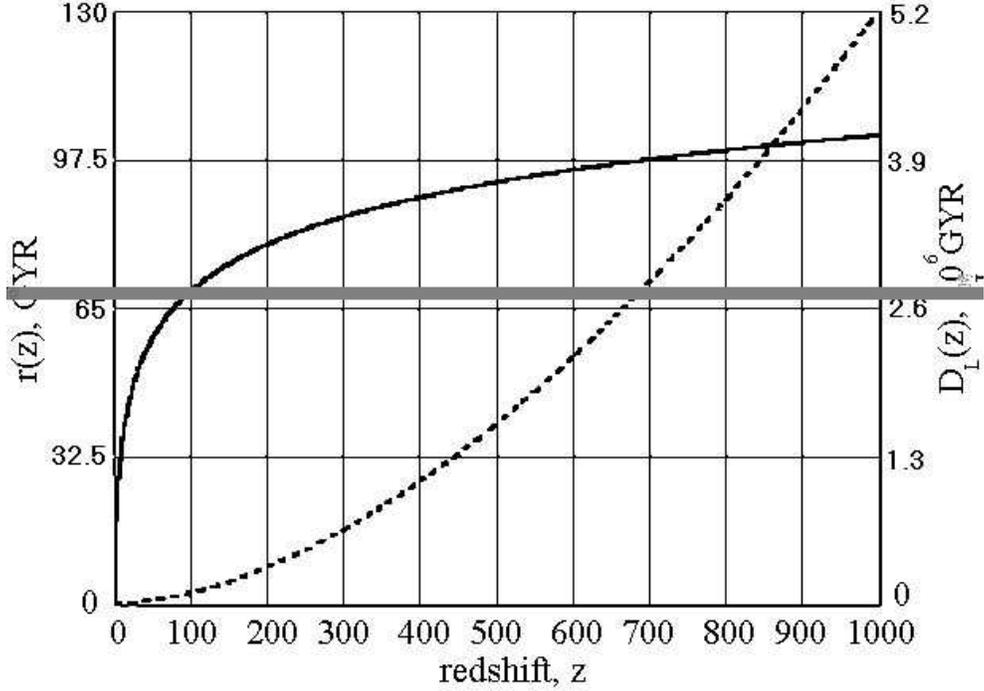}} \caption{The
same functions as on Fig. 6 (all notations are reserved), but for
the huge range of $z$.}
\end{figure}
\par Some other possible cosmological
consequences of an existence of the graviton background were
described in \cite{08,6}. Observations of last years give us
strong evidences for supermassive and compact objects (named now
supermassive black holes) in active and normal galactic nuclei
\cite{65,67,68,612,616}. Massive nuclear "black holes" of $10^{6}
- 10^{9}$ solar masses may be responsible for the energy
production in quasars and active galaxies \cite{65}. In a frame of
this model, an existence of black holes contradicts to the
equivalence principle. It means that these objects should have
another nature; one must remember that we know only that these
objects are supermassive and compact.
\par
There should be two opposite processes of heating and cooling the
graviton background \cite{08} which may have a big impact on
cosmology. Unlike models of expanding universe, in any tired light
model one has a problem of utilization of energy, lost by
radiation of remote objects. In the considered model, a virtual
graviton forms under collision of a photon with a graviton of the
graviton background. It should be massive if an initial graviton
transfers its total momentum to a photon; it follows from the
energy conservation law that its energy $\epsilon^{'}$ must be
equal to $2 \epsilon$ if $\epsilon$ is an initial graviton energy.
In force of the uncertainty relation, one has for a virtual
graviton lifetime $\tau:$ $\tau \leq  \hbar/\epsilon^{'},$ i.e.
for $\epsilon^{'} \sim 10^{-4}~ eV$ it is $\tau \leq 10^{-11}~ s.$
In force of conservation laws for energy, momentum and angular
momentum, a virtual graviton may decay into no less than three
real gravitons. In a case of decay into three gravitons, its
energies should be equal to $\epsilon,~ \epsilon^{''},~ \epsilon
{'''},$ with $\epsilon^{''} + \epsilon {'''}= \epsilon.$ So, after
this decay, two new gravitons with $\epsilon^{''},~ \epsilon {'''}
< \epsilon$ inflow into the graviton background. It is a source of
adjunction of the graviton background.
\par
From another side, an interaction of gravitons of the background
between themselves should lead to the formation of virtual massive
gravitons, too, with energies less than $\epsilon_{min}$ where
$\epsilon_{min}$ is a minimal energy of one graviton of an initial
interacting pair. If gravitons with energies $\epsilon^{''},~
\epsilon {'''}$ wear out a file of collisions with gravitons of
the background, its lifetime increases. In every such a
collision-decay cycle, an average energy of "redundant" gravitons
will double decrease, and its lifetime will double increase. Only
for $\sim 93$ cycles, a lifetime will increase from $10^{-11}~ s$
to $10$ Gyr. Such virtual massive gravitons, with a lifetime
increasing from one collision to another, would duly serve dark
matter particles. Having a zero (or near to zero) initial velocity
relative to the graviton background, the ones will not interact
with matter in any manner excepting usual gravitation. An
ultra-cold gas of such gravitons will condense under influence of
gravitational attraction into "black holes" or other massive
objects. Additionally to it, even in absence of initial
heterogeneity, the one will easy arise in such the gas that would
lead to arising of super compact massive objects, which will be
able to turn out "germs" of "black holes". It is a method "to
cool" the graviton background.
\par
So, the graviton background may turn up "a perpetual engine" of
the universe, pumping energy from any radiation to massive
objects. An equilibrium state of the background will be ensured by
such a temperature $T,$ for which an energy profit of the
background due to an influx of energy from radiation will be equal
to a loss of its energy due to a catch of virtual massive
gravitons with "black holes" or other massive objects. In such the
picture, the chances are that "black holes" would turn out "germs"
of galaxies. After accumulation of a big enough energy by a "black
hole" (to be more exact, by a super-compact massive object) by
means of a catch of virtual massive gravitons, the one would be
absolved from an energy excess in via ejection of matter, from
which stars of galaxy should form. It awaits to understand else in
such the approach how usual matter particles form from virtual
massive gravitons. \par There is a very interesting but
non-researched possibility: due to relative decreasing of an
intensity of graviton pair flux in an internal area of galaxies
(pairs are destructed under collisions with matter particles), the
effective Newton constant may turn out to be running on galactic
scales. It might lead to something like to the modified Newtonian
dynamics (MOND) by  Mordehai Milgrom (about MOND, for example, see
\cite{99}). But to evaluate this effect, one should take into
account a relaxation process for pairs, about which we know
nothing today. It is obvious only that gravity should be stronger
on a galactic periphery. The renormalization group approach to
gravity leads to modifications of the theory of general relativity
on galactic scales \cite{799,899}, and a growth of Newton's
constant at large distances takes place, too. Kepler's third law
receives quantum corrections that may explain the flat rotation
curves of the galaxies.

\section[6]{How to verify the main conjecture of this approach in
a laser experiment on the Earth} I would like to show here (see
\cite{112}) a full realizability at present time of verifying my
basic conjecture about the quantum gravitational nature of
redshifts in a ground-based laser experiment. Of course, many
details of this precision experiment will be in full authority of
experimentalists. \par It was not clear in 1995 how big is a
temperature of the graviton background, and my proposal \cite{111}
to verify the conjecture about the described local quantum
character of redshifts turned out to be very rigid: a laser with
instability of $\sim 10^{-17}$ hasn't appeared after 9 years. But
if $T=2.7 K$, the satellite of main laser line of frequency $\nu$
after passing the delay line will be red-shifted at $\sim 10^{-3}$
eV/h and its position will be fixed (see Fig. 8). It will be
\begin{figure}[th]
\centerline{\includegraphics[width=12.98cm]{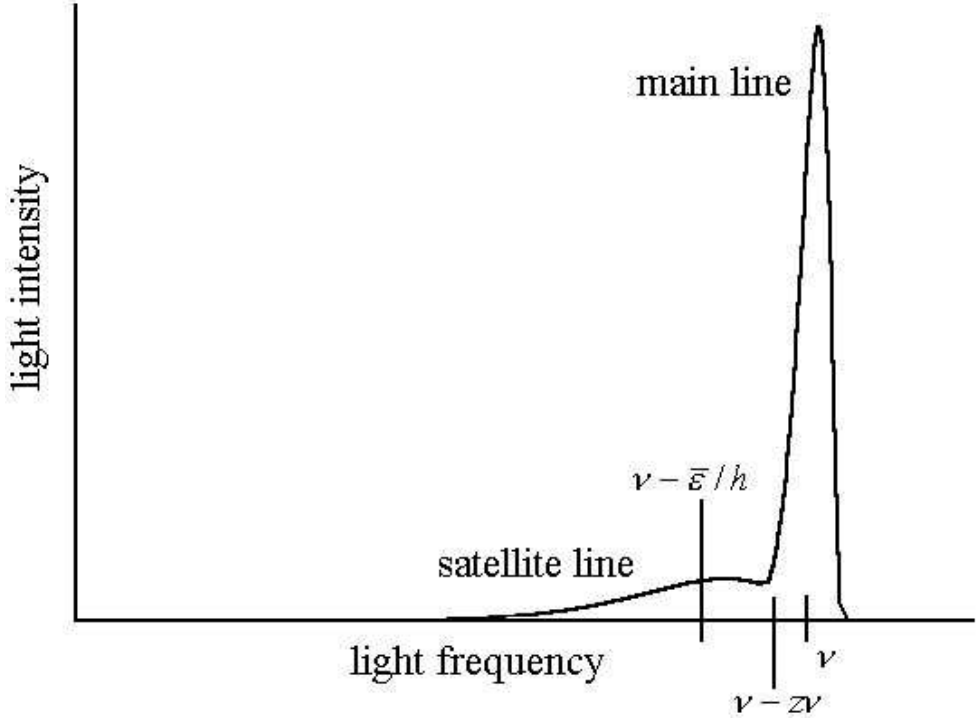}}
\caption{The main line and the expected red-shifted satellite line
of a stable laser radiation spectrum after a delay line.
Satellite's position should be fixed near $\nu -\bar{\epsilon}/h$,
and its intensity should linear rise with a path of photons in a
delay line, $l$. A center-of-mass of both lines is expected to be
approximately near $\nu - z \nu$.}
\end{figure}
caused by the fact that on a very small way in the delay line only
a small part of photons may collide with gravitons of the
background. The rest of them will have unchanged energies. The
center-of-mass of laser radiation spectrum should be shifted
proportionally to a photon path. Then due to the quantum nature of
shifting process, the ratio of satellite's intensity to main
line's intensity should have the order: $$\sim {h\nu \over
\bar{\epsilon}}{H\over c} l,$$ where $l$ is a path of laser
photons in a vacuum tube of delay line. It gives us a possibility
to plan a laser-based experiment to verify the basic conjecture of
this approach with much softer demands to the equipment. An
instability of a laser of a power $P$ must be only $\ll 10^{-3}$
if a photon energy is of $\sim 1~eV$. It will be necessary to
compare intensities of the red-shifted satellite at the very
beginning of the path $l$ and after it. Given a very low
signal-to-noise ratio, one could use a single photon counter to
measure the intensities. When $q$ is a quantum output of a cathode
of the used photomultiplier (a number of photoelectrons is $q$
times smaller than a number of photons falling to the cathode),
$N_{n}$ is a frequency of its noise pulses, and $n$ is a desired
ratio of a signal to noise's standard deviation, then an evaluated
time duration $t$ of data acquisition would have the order:
\begin{equation}
t= {\bar{\epsilon}^{2}c^{2} \over H^{2}} {n^{2}N_{n} \over q^{2}
P^{2} l^{2} }.
\end{equation}
Assuming $n=10,~N_{n}=10^{3}~s^{-1},~ q=0.3, ~P=100~ mW,~ l=100
~m, $ we would have the estimate: $t= 200,000 $ years, that is
unacceptable. But given $P=300~W$, we get: $t \sim 8$ days, that
is acceptable for the experiment of such the potential importance.
Of course, one will rather choose a bigger value of $l$ by a small
laser power forcing a laser beam to whipsaw many times between
mirrors in a delay line - it is a challenge for experimentalists.
\section[7]{Gravity in a frame of non-linear and non-local QED? -
the question only to the Nature} From thermodynamic reasons, it is
assumed here that the graviton background has the same temperature
as the microwave background. Also it follows from the condition of
detail equilibrium, that both backgrounds should have the
Planckian spectra. Composite gravitons will have spin $2$, if
single gravitons have the same spin as photons. The question
arise, of course: how are gravitons and photons connected? Has the
conjecture by Adler et al. \cite{a98,a99} (that a graviton with
spin $2$ is composed with two photons) chances to be true?
Intuitive demur calls forth a huge self-action, photons should be
endued with which if one unifies the main conjecture of this
approach with the one by Adler et al. - but one may get a unified
theory on this way.
\par To verify this combined conjecture in experiment, one would
search for transitions in interstellar gas molecules caused by the
microwave background, with an angular momentum change
corresponding to absorption of spin $2$ particles (photon pairs).
A frequency of such the transitions should correspond to an
equivalent temperature  of the sub-system of these composite
particles $T_{2}=0.5946~ T,$ if $T$ is a temperature  of the
microwave background.
\par
From another side, one might check this conjecture in a laser
experiment, too. Taking two lasers with photon energies $h\nu_{1}$
and $h\nu_{2}$, one may force laser beams to collide on a way $L$
(see Fig. 9). If photons are self-interacting particles, we might
wait that photons with energies $h\nu_{1}-h\nu_{2}$, if $h\nu_{1}
> h\nu_{2}$, would arise after collisions of initial photons.
\begin{figure}[th]
\centerline{\includegraphics[width=12.98cm]{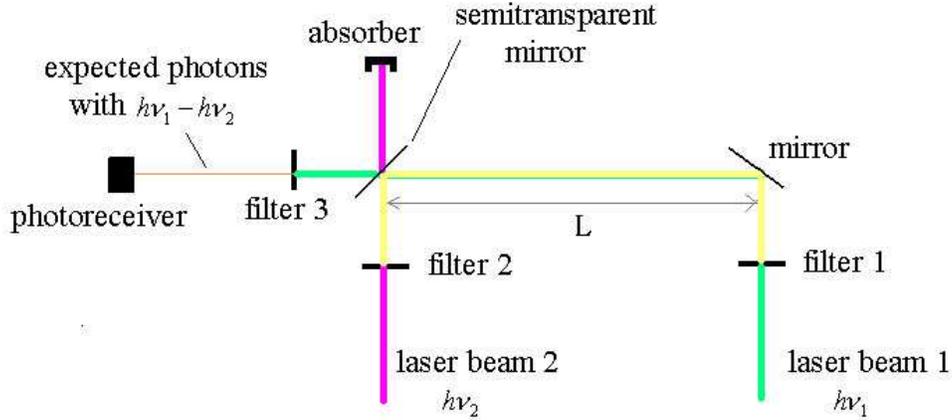}}
\caption{The scheme of laser beam passes. Two laser beams 1 and 2
collide into the area with a length $L$. An expected beam of
photons with energies $h\nu_{1}-h\nu_{2}$ falls to a
photoreceiver.}
\end{figure}
If we {\it assume (only here)} that single gravitons are identical
to photons, it will be necessary to take into account the
following circumstances to calculate an analog of the Hubble
constant for this experiment: an average graviton energy should be
replaced with $h\nu_{2}$, the factor $1/2\pi$ in (8) should be
replaced with $1/\varphi$, where $\varphi$ is a divergence of
laser beam 2, and one must use a quantity $P/S$ instead of $\sigma
T^{4}$ in (8), where $P$ is a laser 2 power and $S$ is a
cross-section of its beam. Together all it means that we should
replace the Hubble constant with its analog for a laser beam
collision, $H_{laser}$:
\begin{equation}
H \rightarrow H_{laser} = {1 \over \varphi} \cdot D \cdot
h\nu_{2}\cdot {P \over S}.
\end{equation}
Taken $\varphi=10^{-4}$, $h\nu_{2} \sim 1~eV$, $P \sim 10~mW$, and
$P/S \sim 10^{3}~W/m^{2}$, that is characterizing a He-Ne laser,
we get the estimate: $H_{laser} \sim 0.06 ~s^{-1}$. Then photons
with energies $h\nu_{1}-h\nu_{2}$ would fall to a photoreceiver
with a frequency which should linearly rise with $L$
(proportionally to ${H_{laser} \over c} \cdot L$), and it would be
of $10^{7}~s^{-1}$ if both lasers have equal powers $\sim 10~mW$,
and $L\sim 1~m$. It is a big enough frequency to give us a
possibility to detect easy a flux of these expected photons in IR
band.
\par I think there is not any sense to try to analyze
theoretically consequences of this conjecture - it will be easier
to verify it experimentally. The Nature may answer the question if
we ask correctly. All that was said in the above sections doesn't
depend on the answer, but it would be very important for our
understanding of known interactions. If this tentative non-linear
vacuum effect exists, it would lead us far beyond standard quantum
electrodynamics to take into account new non-linearities (which
are not connected with electron-positron pair creation) and an
essential impact of such a non-locally born object as the graviton
background.

\section[8]{Conclusion}
It follows from the above consideration that the geometrical
description of gravity should be a good idealization for any pair
of bodies at a big distance by the condition of an "atomic
structure" of matter. This condition cannot be accepted only for
black holes which must interact with gravitons as aggregated
objects. In addition, the equivalence principle is roughly broken
for black holes, if the described quantum mechanism of classical
gravity is realized in the nature. Because attracting bodies are
not initial sources of gravitons, a future theory must be
non-local in this sense to describe gravitons running from
infinity. Non-local models were considered by G.V. Efimov in his
book \cite{a14}. The Le Sage's idea to describe gravity as caused
by running {\it ab extra} particles was criticized by the great
physicist Richard Feynman in his public lectures at Cornell
University \cite{a15}, but the Pioneer 10 anomaly \cite{1},
perhaps, is a good contra argument pro this idea. \par The
described quantum mechanism of classical gravity is obviously
asymmetric relative to the time inversion. By the time inversion,
single gravitons would run against bodies to form pairs after
collisions with bodies. It would lead to replacing a body
attraction with a repulsion. But such the change will do
impossible the graviton pairing. Cosmological models with the
inversion of the time arrow were considered by Sakharov
\cite{a16}. Penrose has noted that a hidden physical law may
determine the time arrow direction \cite{a17}; it will be very
interesting if namely realization in the nature of Newton's law
determines this direction. \par A future theory dealing with
gravitons as usual particles should have a number of features
which are not characterizing any existing model to image the
considered here features of the possible quantum mechanism of
gravity. If this mechanism is realized in the nature, both the
general relativity and quantum mechanics should be modified. Any
divergencies, perhaps, would be not possible in such the model
because of natural smooth cut-offs of the graviton spectrum from
both sides. Gravity at short distances, which are much bigger than
the Planck length, needs to be described only in some unified
manner.

\end{document}